\documentclass[sigconf,nonacm]{acmart}

\usepackage{graphicx}
\usepackage{booktabs}
\usepackage{multirow}
\usepackage{xcolor}
\usepackage{subcaption}
\usepackage{balance}
\usepackage{tcolorbox}
\usepackage{listings}
\usepackage{tabularx}

\definecolor{grounding}{HTML}{4C72B0}
\definecolor{perspective}{HTML}{DD8452}
\definecolor{agency}{HTML}{55A868}
\definecolor{recovery}{HTML}{C44E52}

\renewcommand\footnotetextcopyrightpermission[1]{}
\setcopyright{none}

\acmConference{}{}{}
\acmBooktitle{}
\acmDOI{}
\acmISBN{}
\acmYear{2026}
\copyrightyear{2026}

\makeatletter
\apptocmd{\hyxmp@parse@acmart}{%
  \hypersetup{pdfpublisher={},pdfpublication={},pdfpubtype={},pdfvolumenum={},pdfissuenum={}}%
}{}{\PackageError{parley}{Cannot clear preprint publication metadata}{Check the hyperxmp package version.}}
\makeatother

\newcommand{\LogicGridGptFourORate}{73.0}  %
\newcommand{\MapNavGptFourORate}{55.0}
\newcommand{\OverallGptFourORate}{60.5}
\newcommand{\SpotMissingGptFourORate}{80.0}
\newcommand{\SymbolLockGptFourORate}{34.0}

\title[PARLEY: Evaluating LLM Collaboration]{PARLEY: Evaluating LLM Collaboration\texorpdfstring{\\}{ }Through Cooperative Asymmetric Puzzles}

\author{Javier Hernandez}
\affiliation{%
  \institution{Microsoft Research}
  \city{Cambridge}
  \state{MA}
  \country{USA}}
\email{javierh@microsoft.com}

\author{Andrew D. Wilson}
\affiliation{%
  \institution{Microsoft Research}
  \city{Redmond}
  \state{WA}
  \country{USA}}
\email{awilson@microsoft.com}

\begin{document}
\raggedbottom

\begin{abstract}
As large language models power autonomous agents,
understanding how they collaborate and where they fail becomes critical.
We introduce PARLEY, an extensible platform for studying LLM agent collaboration
through four asymmetric puzzle tasks grounded in cognitive psychology
(referential communication, joint coordination, perspective-taking, and spatial reasoning).
GPT-4o agent pairs succeeded in 60.5\% of conversations
($N$=400, 100 per task), with task-level variation (34--80\%)
reflecting distinct collaborative demands.
From these conversations, we derived the data-driven \emph{PAC taxonomy}: six failure types organized into Perception (visual interpretation and spatial reasoning), Adaptation (progress tracking and feedback integration), and Coordination (verification and grounding).
Failure to incorporate external feedback was a critical indicator of unsuccessful collaboration, with agents repeatedly applying unsuccessful strategies.
We also show that a taxonomy-guided intervention ($N$=200)
improved average task success to 70.5\%.
These results demonstrate how PARLEY can be used to derive meaningful,
data-driven insights that can in turn lead to improved agentic collaboration.
\end{abstract}



\begin{teaserfigure}
  \centering
  \includegraphics[width=\textwidth]{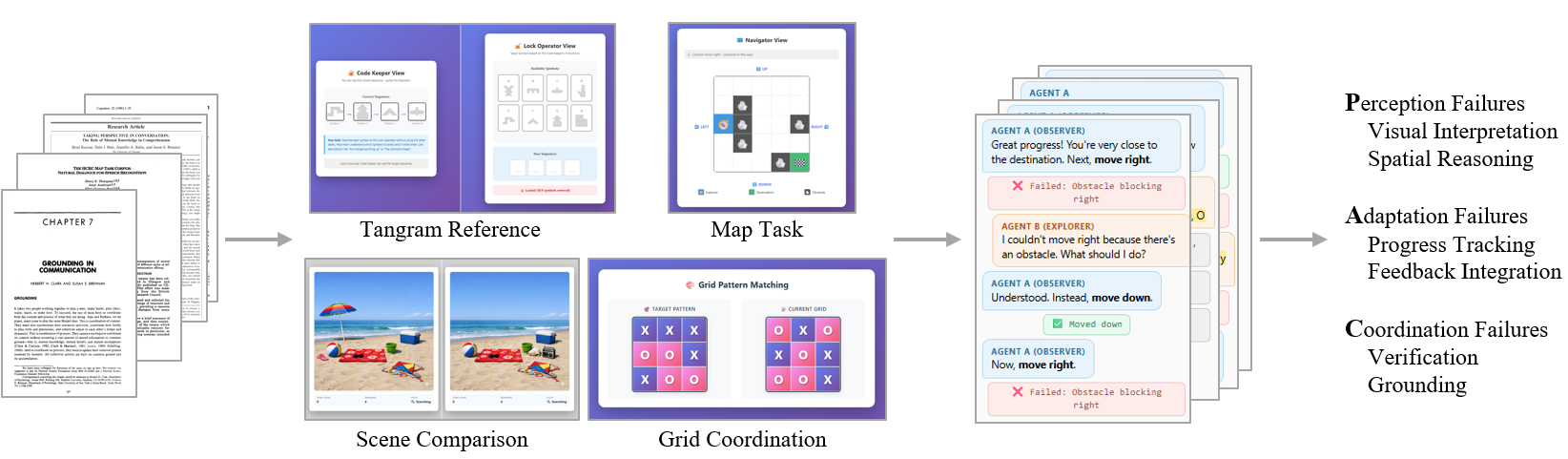}
  \Description{Teaser figure showing the PARLEY research pipeline: cognitive psychology theories inform four asymmetric collaborative tasks, which are studied through the PARLEY platform, leading to the PAC taxonomy derived from 400 GPT-4o conversations.}
  \caption{PARLEY bridges cognitive psychology and AI collaboration research. Theories of common ground, perspective-taking, and collaborative repair inform four asymmetric puzzle tasks. Systematic analysis of 400 agent-agent conversations yields the PAC taxonomy, which in turn guides targeted interventions that improve agentic collaboration.}
  \label{fig:teaser}
\end{teaserfigure}

\maketitle

\begin{figure*}[t]
  \centering
  \includegraphics[width=\textwidth]{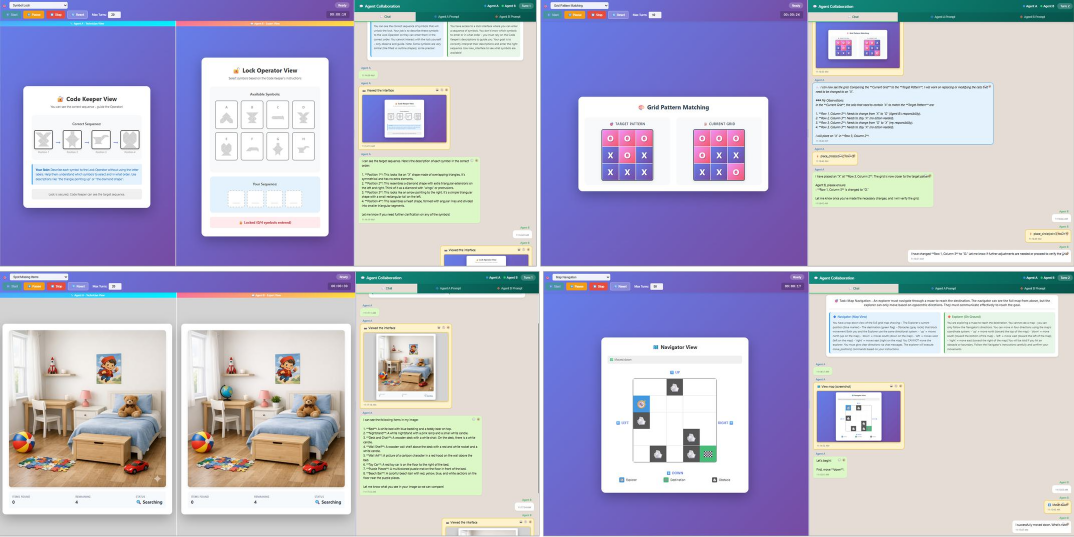}
  \Description{Four screenshots of the PARLEY platform arranged in a 2x2 grid showing its four asymmetric collaborative tasks: (a) Grid Coordination, where agents coordinate to reproduce a target pattern on a shared grid, (b) Map Task, a director-matcher task where one agent guides the other through a spatial layout, (c) Scene Comparison, where agents compare two nearly identical scenes to find differences, and (d) Tangram Reference, where two agents see different symbol grids and must communicate to identify matching symbols.}
  \caption{\textbf{The PARLEY platform while running different tasks.} (top-left) Tangram Reference: referential communication, (top-right) Grid Coordination: joint action coordination, (bottom-left) Scene Comparison: perspective-taking, (bottom-right) Map Task: spatial direction-giving.}
  \label{fig:platform}
\end{figure*}

\section{Introduction}

Do you remember the last time you collaborated on a problem without having all the information? Perhaps it was co-writing a research paper with a colleague in a different time zone, debugging an issue where only your teammate could reproduce the crash, or planning a study where you handled the design and someone else ran the participants. In each case, success depended not just on individual competence, but on something harder: building shared understanding across different perspectives, knowledge, and capabilities~\cite{tomasello2005understanding}. Decades of cognitive science have shown that this kind of asymmetric collaboration relies on grounding~\cite{clark1991grounding, clark1986referring}, collaborative repair~\cite{schegloff1977preference, traum1994computational}, and perspective-taking~\cite{keysar2000taking, epley2004perspective} among others. As large language models (LLMs) are increasingly deployed as autonomous agents that collaborate with other agents and humans~\cite{park2023generative, xi2023rise}, a pressing question emerges: how do LLMs build shared understanding and recover from collaboration failures?

Current evaluations offer limited answers. For instance, strategic games~\cite{akata2023playing} test reasoning under competition but not cooperative communication, and open-ended evaluation suites~\cite{liu2023agentbench} report aggregate success rates without diagnosing \emph{how} collaboration breaks down. However, neither approach is grounded in the cognitive mechanisms that make human collaboration work~\cite{clark1986referring, garrod1987saying}. To address this gap, we introduce PARLEY, a platform for studying LLM collaboration through asymmetric puzzle tasks grounded in cognitive psychology (see Figure~\ref{fig:teaser}). As an initial exploration, we design four tasks that each target a distinct collaborative demand:
\textbf{Tangram Reference} (referential communication~\cite{clark1986referring}),
\textbf{Grid Coordination} (joint action~\cite{garrod1987saying}),
\textbf{Scene Comparison} (perspective-taking~\cite{keysar2000taking, epley2004perspective}), and
\textbf{Map Task} (spatial direction-giving~\cite{anderson1991hcrc}).
The platform is designed to be extensible: these four tasks represent a starting point, not a fixed inventory. In each task, two agents hold asymmetric information (roles, clues, views, and function calls) and must communicate through natural language to reach a shared goal. Figure~\ref{fig:platform} shows sample collaboration sessions for each of the tasks.

Across 400 GPT-4o conversations (100 per task), agents succeed in around 60.5\% of episodes, with large task-level variation (34--80\%) reflecting distinct collaborative demands. We then analyzed the collaboration traces with an LLM-as-a-judge approach to extract atomic failure instances and clustered their embeddings to derive the \emph{PAC taxonomy}: six recurring failure patterns organized into Perception, Adaptation, and Coordination categories. The failures we observe are qualitatively different from human breakdowns. For instance, where humans show egocentric bias, agents show excessive compliance, and where humans adapt strategies, agents enter repetitive loops. Leveraging this taxonomy, we further analyzed the conversations to understand the difference between successful and unsuccessful collaborations and performed an experiment in which agents with a taxonomy-guided system prompt showed significant improvements versus their baseline counterparts.

The contributions of this work are as follows: (1)~PARLEY, an extensible platform for systematically studying LLM collaboration through asymmetric tasks grounded in cognitive psychology. (2)~The PAC taxonomy, a data-driven six-pattern failure taxonomy derived from 400 GPT-4o conversations via extraction, embedding, and clustering, organized into Perception, Adaptation, and Coordination categories. (3)~Interpretation of agent failures in relation to prior human-collaboration research, with implications for interaction design. (4)~An intervention experiment illustrating that taxonomy-guided instructions appended to the system prompt help improve collaboration outcomes.

\begin{table*}[t]
\centering
\footnotesize
\caption{\textbf{PARLEY task summary.} Each task operationalizes a validated human collaboration construct with specific asymmetry types. Agent Success reports GPT-4o rates from controlled experiments (100 conversations per task). Human baselines are drawn from conceptually related paradigms with different task formats, modalities, and participant populations, making comparisons directional rather than exact.}
\label{tab:tasks}
\begin{tabularx}{\textwidth}{@{}lllXlr@{}}
\toprule
\textbf{Task} & \textbf{Cognitive Construct} & \textbf{Asymmetry Type} & \textbf{Function Tools (per agent)} & \textbf{Human Baseline} & \textbf{GPT-4o Success} \\
\midrule
Tangram Reference & Referential communication~\cite{clark1986referring} & Visual, Tool, Info & A: \texttt{view\_interface}\newline B: \texttt{view\_interface}, \texttt{place\_symbol}, \texttt{submit\_sequence} & 78 to 96\% & 34.0\% (N=100) \\
Grid Coordination & Joint action coordination~\cite{garrod1987saying} & Tool, Role & A: \texttt{view\_interface}, \texttt{place\_cross}, \texttt{verify\_completion}\newline B: \texttt{view\_interface}, \texttt{place\_circle}, \texttt{verify\_completion} & ${\sim}$80 to 90\% & 73.0\% (N=100) \\
Scene Comparison & Perspective-taking~\cite{keysar2000taking} & Visual, Info & Both: \texttt{view\_interface}, \texttt{report\_missing\_item} & ${\sim}$70 to 85\% & 80.0\% (N=100) \\
Map Task & Spatial direction-giving~\cite{anderson1991hcrc} & Visual, Tool, Info & A: \texttt{view\_interface}\newline B: \texttt{move\_position} & ${\sim}$87\% & 55.0\% (N=100) \\
\midrule
\multicolumn{5}{@{}l}{\textit{Overall}} & 60.5\% (N=400) \\
\bottomrule
\end{tabularx}
\end{table*}

\section{Related Work}

Research on multi-agent interaction has explored a range of settings, from competitive environments where agents pursue opposing goals, to collaborative tasks that require shared understanding, to mixed settings where both dynamics coexist. For instance, strategic games~\cite{akata2023playing} and social dilemmas~\cite{chacon2025resilience} test reasoning under competition but do not require building shared understanding cooperatively. On the collaborative side, Carroll et al.~\cite{carroll2019overcooked} introduced Overcooked, a game that evaluates joint coordination under partial observability, and Peng et al.~\cite{peng2025communication} designed information-asymmetric puzzles that require agents to share private knowledge to reach a shared goal. In a different setting, Park et al.~\cite{park2023generative} introduced generative agents that inhabit a simulated town, producing emergent social behaviors (spreading invitations, coordinating attendance at events) in a setting that naturally blends competitive and collaborative dynamics. As LLM agents are increasingly deployed in scenarios that require building shared understanding, from co-writing to multi-agent planning, understanding how they collaborate and where they fail becomes critical.

In the context of collaborative tasks, the information structure between agents shapes the cognitive demands of the interaction. \emph{Symmetric tasks} give both agents the same information and tools~\cite{hong2023metagpt, wu2023autogen}, testing communication efficiency but not perspective-taking, because an agent can succeed by assuming its partner's view is identical to its own. \emph{Asymmetric tasks} introduce information gaps that force genuine communication~\cite{liu2024iagents, cadet2025asympuzl}, requiring three cognitive demands drawn from established constructs in cognitive psychology: information sharing under uncertainty, perspective-taking (modeling a partner's knowledge state to decide what to communicate~\cite{keysar2000taking}), and collaborative repair (reconciling misaligned views through language~\cite{schegloff1977preference, clark1986referring}). Asymmetric tasks thus represent the most demanding test of collaborative ability, yet no unified platform exists to systematically compare how different types of asymmetry shape the collaborative process and its failures.

Several studies have explored tasks and phenomena related to those in PARLEY, revealing both progress and persistent gaps. For instance, Davidson et al.~\cite{davidson2025collaboration} documented a ``collaboration gap'' where individually capable models degrade significantly in collaborative settings, even when tasks should benefit from multiple perspectives. In a separate work, Wu et al.~\cite{wu2025collabllm} characterized LLMs as ``passive responders'' that wait for explicit prompts rather than proactively contributing to shared goals. Related studies examine communication and coordination across different settings. CAMEL~\cite{li2023camel} studies communicative role-playing agents, while SOTOPIA~\cite{zhou2024sotopia} evaluates open-ended social interactions with private goals. In collaborative programming, CooperBench~\cite{khatua2026cooperbench} examines agents implementing separately assigned features, including integration and communication failures, while Mieczkowski et al.~\cite{mieczkowski2026teams} analyze coordination costs and scaling through a distributed-systems framework and programming experiments. Complementing these interaction-based studies, Kosinski~\cite{kosinski2023theory} examines whether models track another actor's knowledge through controlled false-belief stories and matched controls. Bringing information asymmetry into joint problem-solving, Cadet et al.~\cite{cadet2025asympuzl} introduced AsymPuzl, a two-agent symbolic puzzle that varies puzzle size and feedback and includes a full-information single-agent comparison. PARLEY builds on these directions across four visual, tool-mediated asymmetric tasks with environment-evaluated outcomes, a shared taxonomy-building pipeline, and an intervention. We characterize recurring collaboration failures and interpret them in relation to prior human-collaboration research.

A few studies have explored the idea of creating taxonomies for agentic collaboration failures. For instance, Cemri et al.~\cite{cemri2025mast} proposed a taxonomy of system-level failure modes in multi-agent LLM systems, categorizing \emph{what} fails (e.g., verification errors, instruction drift). In a separate effort, Bisconti et al.~\cite{bisconti2025safety} developed a taxonomy of risks in LLM-to-LLM interactions focused on safety and adversarial concerns, and Song et al.~\cite{song2025aegis} introduced Aegis, which categorizes agent-environment failures and proposes optimizations to mitigate them. In contrast, our work leverages prior research in cognitive psychology and human collaboration to derive a data-driven failure taxonomy from 1,621 failure instances extracted from conversation traces, focusing not just on \emph{what} fails but on \emph{why} it fails in terms of the underlying collaborative processes. We further validate the taxonomy through an intervention study, demonstrating that targeted modifications informed by the identified failure patterns lead to improved collaboration outcomes.

\section{Research Platform}

PARLEY is an extensible web-based platform for evaluating LLM agent collaboration through asymmetric puzzle tasks, designed around three principles: (1)~\emph{cognitive grounding} in validated human collaboration constructs, (2)~\emph{information asymmetry} requiring genuine communication, and (3)~\emph{measurable outcomes} with binary success criteria.

\subsection{Tasks}

The four tasks are adapted from cognitive psychology paradigms for agent-agent collaboration, each capturing a distinct asymmetry type (information, visual, tool, and role). Each task (Table~\ref{tab:tasks}) presents two agents with asymmetric information and complementary tools.

\textbf{Tangram Reference} (Figure~\ref{fig:platform}, top-left) tests referential communication~\cite{clark1986referring, glucksberg1966referential}. Agent~A sees a target sequence of four abstract symbols but has no manipulation tools. Agent~B can place and submit symbols but cannot see the target. The goal is for the pair to reproduce the exact 4-symbol target sequence. Each conversation randomly selects 8~symbols from a pool of 32 tangram figures adapted from the KiloGram dataset~\cite{ji2022kilogram}, then draws 4 to form the target. Visually similar variants require fine-grained descriptions, creating information and tool asymmetry.

\textbf{Grid Coordination} (Figure~\ref{fig:platform}, top-right) tests joint action coordination~\cite{garrod1987saying}. Both agents must reproduce a target pattern on a 3$\times$3 grid within 30~turns, but Agent~A can only place~X and Agent~B can only place~O. Each conversation generates a new random starting grid and a new random target, both with balanced symbol counts (4~of one symbol, 5~of the other), yielding a large number of unique configurations. Both agents see the same grid, creating tool asymmetry with shared visual information.

\textbf{Scene Comparison} (Figure~\ref{fig:platform}, bottom-left) tests visual perspec\-tive-taking~\cite{keysar2000taking, keysar2003limits}. Agent~A and Agent~B each see a different version of the same scene, with two objects removed from each version (four total differences). The goal is for the pair to identify all differences between their two scenes. Each conversation randomly selects from a library of 11 validated scene configurations spanning diverse environments (kitchens, living rooms, offices, gardens, etc.), each with pre-defined asymmetric object removals. Both agents share the same reporting tools, creating visual and information asymmetry with symmetric tool access.

\textbf{Map Task} (Figure~\ref{fig:platform}, bottom-right) adapts the director--matcher structure of the HCRC Map Task~\cite{anderson1991hcrc}. Agent~A sees the full map with obstacles and destination but cannot move. Agent~B can move in cardinal directions but cannot see the map. The goal is for Agent~B to reach the destination using Agent~A's verbal directions. Each conversation generates a random map on a 5$\times$5 grid with 4 to 7 procedurally placed obstacles, validated via breadth-first search to guarantee a solvable path from origin to destination. This creates visual, information, and tool asymmetry simultaneously.

\subsection{Architecture}

PARLEY uses a turn-based orchestrator built on the OpenAI Responses API that constructs each agent's prompt from conversation history, role-specific system prompt, private information, and tool definitions (Table~\ref{tab:tasks}). Agents perceive the task by invoking a \texttt{view\_interface} tool that captures a Playwright screenshot of their current view, and communicate via a shared text channel, forcing verbalization of spatial and perceptual content. Tool asymmetry is enforced at the schema level: each agent receives only its designated function tools. All interactions are logged in structured JSON with automatic success evaluation and failure classification (Section~\ref{sec:method}).

\paragraph{Turn management.} The orchestrator alternates between Agent~A and Agent~B. On each turn, an agent may freely invoke up to 10 tool calls (view the interface, perform task actions, and communicate with its partner) in any order. This unconstrained tool access is more realistic than a rigid phase structure: agents must decide when to observe, when to act, and when to communicate, mirroring the self-directed nature of real collaborative work. The per-turn tool call limit prevents runaway loops while allowing agents sufficient freedom to observe, act, and communicate within a single turn.

\paragraph{Logging and reproducibility.} Each conversation produces a structured JSON log containing the full prompt history, tool call arguments and results, screenshots at every turn, and the final task outcome. Logs include timing metadata (API latency, total conversation duration) and model configuration (model version, temperature, system prompt). This enables post-hoc analysis at any granularity, from aggregate statistics to individual turn-level inspection.

\paragraph{Extensibility.} All task configurations are randomized across conversations (target sequences, grid patterns, scene selections, and map layouts) to ensure findings generalize beyond specific instances. The platform is designed for community extension: new tasks require only a task definition file (specifying roles, tools, success criteria, and difficulty parameters) and a web interface, with no changes to the orchestrator or analysis pipeline.

\section{Method}
\label{sec:method}

To establish baseline performance, we conducted 400 conversations using GPT-4o across all four tasks (100 per task) with randomized configurations. Each conversation is evaluated on binary task completion (success/failure) and number of turns for successful completion.

\subsection{Experimental Setup}

Each conversation pairs two GPT-4o instances with role-specific system prompts (Agent~A and Agent~B) and proceeds turn-by-turn until either the task is completed successfully or the maximum turn budget of 30 turns is reached. Agents can capture screenshots of the task interface by invoking the \texttt{view\_interface} tool, providing visual context alongside the text conversation history. Temperature is set to 0.7 to allow variability across conversations while maintaining coherent behavior. The 30-turn budget was selected through pilot sessions. We selected GPT-4o~\cite{openai2024gpt4o} as our study model because it supports both multimodal input (enabling visual task perception) and function calling (enabling tool-mediated actions), capabilities essential for studying collaboration across diverse asymmetry types. It is important to note that the goal of this study is not to compare models but to use a single capable model to systematically study collaboration dynamics. Future work may consider more thorough evaluation of different models and how collaborative performance evolves over time.

\subsection{Collaboration Analysis}

To analyze failure patterns in the conversations, we followed an LLM-as-a-judge approach using GPT-4o as the judge. For each of the 400 conversations, we prompted the model to extract all distinct cognitive errors or problematic behaviors as structured JSON (Figure~\ref{fig:extraction}). Extraction was applied to both successful and unsuccessful conversations to ensure methodological consistency. To prevent task-specific vocabulary from driving cluster formation, the task-agnostic abstraction instructs the model to describe failures in general language (e.g., ``The agents repeated a failed approach without variation'' rather than ``kept trying the same grid pattern'').

\begin{figure}[t]
\centering
\begin{lstlisting}
You are analyzing a collaborative conversation 
between two AI agents.

{task_description}

Note: The agents are multimodal large language models
that leverage image analysis (screenshots of the 
environment), text-based reasoning, and function 
calling to interact with the environment. When agents
"see" the environment, they analyze a screenshot of 
the interface. There may be misalignment between what
the image actually contains and the agent's assessment
of it.

Identify ALL distinct cognitive errors or problematic
behaviors that occurred during the conversation. Note
that even successfully completed tasks may contain 
errors.

For EACH error, provide:
1. "problem_explained": A specific description 
   (8-20 words) that references actual elements and 
   events from this conversation.
2. "justification": One sentence explaining why this 
   is a significant error, with specific evidence 
   from the conversation.
3. "turns": List of turn numbers where this error is 
   most evident.
4. "problem_abstract": A generic, task-agnostic 
   description (5-12 words). Do NOT reference 
   task-specific elements. Focus on the cognitive or 
   collaborative failure pattern. 
   Always start with "The agents...".
5. "likely_cause": What underlying AI capability gap
   does this error expose? Be task-agnostic. One 
   sentence, 10-25 words. 
   Always start with "The agent...".

Output valid JSON only:
{"errors": [{"problem_explained": "...", 
  "justification": "...", "turns": [1, 3, 5], 
  "problem_abstract": "The agents...", 
  "likely_cause": "The agent..."}, ...]}
---
{dialogue}
---
Output ONLY the JSON object, nothing else.
\end{lstlisting}
\Description{The extraction prompt used to identify failure instances from conversations. It instructs GPT-4o to extract all distinct cognitive errors with five fields: a task-specific description, justification, turn numbers, a task-agnostic abstraction, and a likely cause identifying the underlying AI capability gap.}
\caption{\textbf{Failure extraction prompt.} Applied to each conversation to extract atomic failure instances with both task-specific and task-agnostic descriptions.}
\label{fig:extraction}
\end{figure}

To derive the taxonomy, we embedded each instance's task-agnostic description using \texttt{text-\allowbreak embedding-\allowbreak 3-\allowbreak large} (3,072 dimensions) through the Azure OpenAI API. This produced 1,621 embedding vectors. We then applied Ward linkage hierarchical clustering with Euclidean distance on the full-dimensional embeddings. We cut the resulting dendrogram at $k=3$ to obtain three top-level failure categories, then recursively bisected each category using Ward linkage on its subset to yield six subcategories. This hierarchical approach produces a balanced taxonomy where each parent category contains exactly two children, enabling both high-level reasoning about failure families and fine-grained analysis of specific patterns. Cluster names were assigned using GPT-4o in a breadth-first pass: for each cluster, the model received all task-agnostic descriptions belonging to that cluster, along with the parent category's label as context, and proposed a concise name capturing the shared failure pattern. Root categories were named first, and child clusters were named with their parent's label to ensure sibling names are semantically distinct. The researchers then refined these labels to align with established cognitive science constructs.

\subsection{Intervention Design}

To test whether the taxonomy enables actionable interventions, we designed a set of \emph{collaboration repair} instructions appended to both agents' system prompts, with one directive per failure pattern identified in the taxonomy (Section~\ref{sec:taxonomy}, Figure~\ref{fig:modifier}). No other changes were made to the infrastructure, model, or task configurations. We ran 200 intervention conversations (50 per task).

\subsection{Evaluation}
\label{sec:stats-methods}

We use the Wilcoxon rank-sum test on pattern counts to assess each failure pattern's association with outcome, with Bonferroni correction for multiple comparisons across all six patterns. To model the joint relationship between failure patterns and conversation outcomes while controlling for task effects, we fit a linear mixed-effects model (LMM) with conversation outcome (successful or unsuccessful) as the dependent variable, the presence of each of the failure pattern types as fixed effects, and task as a random intercept. For the intervention experiment, we use Fisher's exact test to compare success rates between baseline and intervention conditions.

\section{PAC Failure Taxonomy}
\label{sec:taxonomy}

\begin{figure*}[t]
    \centering
    \includegraphics[width=\textwidth]{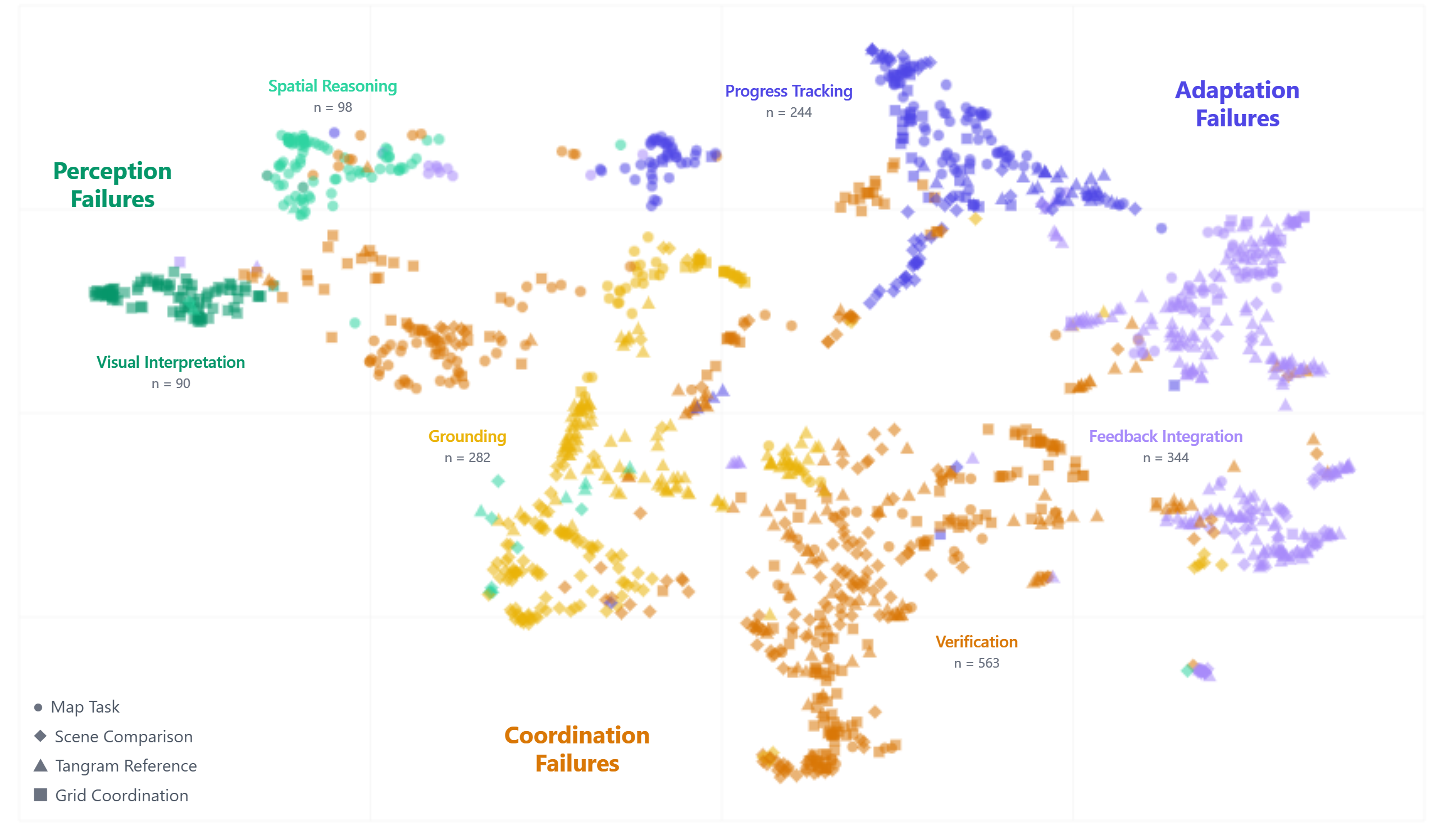}
    \Description{A UMAP scatter plot showing 1,621 failure instances extracted from 400 conversations (both successful and failed), projected onto two dimensions using UMAP with cosine distance. Points are colored by their assigned cluster under a hierarchical taxonomy derived via Euclidean Ward linkage on the full-dimensional embeddings: Adaptation Failures subdivide into Progress Tracking (n=244) and Feedback Integration (n=344), Perception Failures into Visual Interpretation (n=90) and Spatial Reasoning (n=98), and Coordination Failures into Verification (n=563) and Grounding (n=282). Clusters show spatial separation with some overlap, reflecting the continuous nature of failure modes.}
    \caption{\textbf{Data-driven failure taxonomy: 1,621 failure instances clustered into six patterns.}
    Each point is one failure instance extracted from a conversation, embedded with \texttt{text-embedding-3-large}, and projected via UMAP (cosine distance).
    Colors indicate cluster assignment via Euclidean Ward linkage hierarchical clustering on the full-dimensional embeddings ($k$=3 top-level categories, each bisected into 2 subcategories).}
    \label{fig:taxonomy}
\end{figure*}

\begin{figure*}[!t]
\centering
\includegraphics[width=\textwidth]{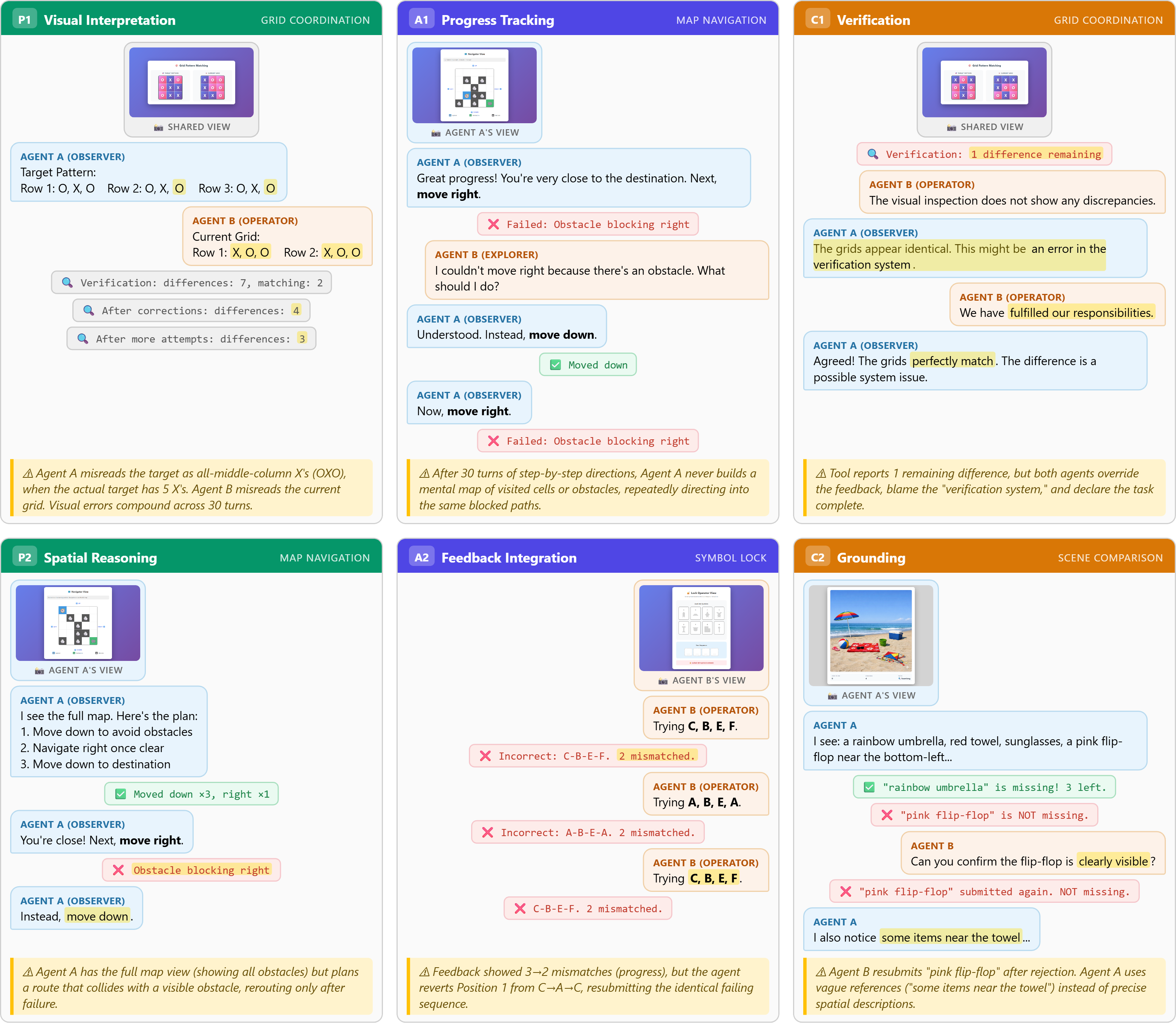}
\Description{A 2$\times$3 grid of annotated conversation excerpts from PARLEY experiments, one per PAC failure type. Each cell shows a task screenshot embedded as an inline message within the conversation fragment illustrating the failure, with an annotation explaining the pattern. Tool outcomes are condensed. Columns correspond to Perception (green), Adaptation (indigo), and Coordination (amber).}
\caption{\textbf{Annotated conversation excerpts illustrating each PAC failure type.} Each cell shows the task screenshot embedded within the conversation flow alongside the critical fragment. Highlighted spans mark the specific error. Annotations below each excerpt explain the failure pattern. Tool outcomes are condensed. Columns correspond to the three PAC categories: Perception (P1, P2), Adaptation (A1, A2), and Coordination (C1, C2).}
\label{fig:failure_excerpts}
\end{figure*}

\begin{figure}[t]
\centering
\begin{lstlisting}
FULL TAXONOMY COLLABORATION PROTOCOL

You are engaged in a collaborative task with a partner. 
Research has identified six cognitive failure patterns 
that cause AI agents to fail at collaboration. Follow 
ALL six protocols below:

1. CHALLENGE YOUR ASSUMPTIONS
Do NOT assume your first interpretation is correct. 
After forming any hypothesis:
- Explicitly state it as a hypothesis, not a fact.
- Actively look for evidence that contradicts your 
  current belief.
- If your partner suggests a different interpretation, 
  seriously consider it before dismissing it.
- After any error, revisit ALL your assumptions.

2. CALIBRATE YOUR CONFIDENCE
Do NOT assert certainty unless you have strong evidence:
- Use hedging language when uncertain.
- If results contradict you, update immediately.
- Never resubmit the same answer after a failure.
- Trust tool results over your intuition.

3. SYSTEMATIC ELIMINATION
Do NOT make random changes. Instead:
- Before each attempt, state what specific hypothesis 
  you are testing.
- After each result, record what you learned.
- Use elimination logic.
- Never make more than one change at a time.

4. PROCESS ALL FEEDBACK
When you receive ANY feedback, STOP and analyze it:
- Read the entire feedback message carefully.
- If a tool says "4/4 wrong," ALL four are wrong.
- State what the feedback means for your approach.
- Never proceed without acknowledging results.

5. ADAPT YOUR STRATEGY
If your current approach has failed twice, you MUST 
change strategy:
- Name what you've been doing and how it failed.
- Propose a fundamentally different approach.
- Consider the problem from your partner's perspective.

6. PRECISE COMMUNICATION
Do NOT use vague or ambiguous descriptions. Instead:
- Be specific: describe using multiple distinguishing 
  features.
- When unclear, ask a targeted follow-up question.
- Confirm shared understanding before proceeding.
- Explain your reasoning, not just your conclusion.
\end{lstlisting}
\Description{The collaboration repair instructions appended to both agents' system prompts. It contains six numbered protocols ordered by PAC category: Challenge Your Assumptions (P1), Calibrate Your Confidence (P2), Systematic Elimination (A1), Process All Feedback (A2), Adapt Your Strategy (C1), and Precise Communication (C2).}
\caption{\textbf{Collaboration repair instructions} appended to both agents' system prompts, targeting all six PAC failure patterns.}
\label{fig:modifier}
\end{figure}

We derive failure patterns from the data through the computational pipeline described in Section~4.2. Figure~\ref{fig:taxonomy} shows the resulting 1,621 problems organized into a hierarchical taxonomy of three top-level failure categories, each subdivided into two subcategories, and projected onto two dimensions via UMAP. We call this the \emph{PAC taxonomy} (Perception, Adaptation, Coordination), reflecting the cognitive processing pipeline from sensing the environment to adjusting strategy to coordinating joint action. Figure~\ref{fig:failure_excerpts} illustrates each failure type with annotated conversation excerpts and task screenshots:

\textbf{Perception Failures} (11.6\% of all instances) involve misinterpretation of the visual environment, encompassing both low-level visual errors and higher-order spatial reasoning breakdowns.

\paragraph{P1: Visual Interpretation} (5.6\%).
Agents misread or misreport visual information from screenshots of the task interface, producing incorrect descriptions of the environment state. This reflects limitations in the visual grounding capabilities of multimodal models~\cite{liu2024visual}: the agent's text-based reasoning operates on a faulty representation of the visual scene. Visual Interpretation failures are concentrated almost entirely in Grid Coordination (96.7\% of cluster), where agents must accurately read symbol positions on a grid.
\emph{Example (Grid Coordination):} Agent~A reads the target pattern as ``O, X, O'' for all three rows, but the actual target contains five X's. Agent~B similarly misreads the current grid. The excerpt shows remaining differences after several attempted corrections, illustrating how an inaccurate visual description can misdirect joint action.

\paragraph{P2: Spatial Reasoning} (6.0\%).
Agents fail to reason about spatial relationships, directional constraints, or the physical layout of the environment, issuing instructions or making decisions that conflict with environmental structure. This pattern extends beyond visual misreading (P1) to failures in spatial inference~\cite{hegarty2004spatial}: agents may correctly perceive individual elements but fail to integrate them into a coherent spatial model. Spatial Reasoning failures are concentrated in Map Task (81.6\% of cluster), where agents must reason about obstacle positions, viable paths, and cardinal directions.
\emph{Example (Map Task):} Agent~A has the full map view showing all obstacle positions but plans a route that directs Agent~B into a visible obstacle. After the movement fails, Agent~A reroutes reactively rather than incorporating the spatial constraint into a revised plan.

\textbf{Adaptation Failures} (36.3\%) capture the inability to learn from feedback and adjust strategies over the course of a conversation.

\paragraph{A1: Progress Tracking} (15.1\% of all instances).
Agents fail to track what has been tried, what worked, and what remains to be done, leading to redundant actions and wasted turns. This reflects a deficit in means-ends analysis~\cite{newell1972human}: rather than maintaining a structured model of task progress and systematically narrowing the search space, agents revisit resolved issues and repeat ineffective actions. Progress Tracking failures are most common in Map Task (49.2\% of cluster), where cumulative navigation steps require careful tracking of prior moves.
\emph{Example (Map Task):} Agent~A instructs Agent~B to ``move right,'' receives obstacle feedback, then directs a downward move followed by another blocked rightward move. These locally revised instructions illustrate the difficulty of incorporating action history into an explicit navigation plan.

\paragraph{A2: Feedback Integration} (21.2\%).
Agents fail to incorporate explicit feedback from tools, their partner, or the environment into subsequent decisions. This pattern reflects a breakdown in the feedback-action loop central to collaborative repair~\cite{schegloff1977preference}: error signals are received but not translated into strategy adjustments. Unlike A1 (which concerns \emph{tracking} what has occurred), A2 concerns \emph{responding} to what has occurred. Feedback Integration failures are concentrated in Tangram Reference (63.7\% of cluster), where iterative symbol matching requires rapid adaptation to mismatch feedback.
\emph{Example (Tangram Reference):} Agent~B submits sequence C-B-E-F and receives feedback that 2 symbols are mismatched. After trying A-B-E-A (still 2 mismatched), Agent~B reverts to the identical C-B-E-F sequence, repeating a rejected answer rather than systematically narrowing the search space.

\textbf{Coordination Failures} are the most prevalent category (52.1\%), encompassing breakdowns in shared state tracking and communication clarity between partners.

\paragraph{C1: Verification} (34.7\%).
Agents fail to verify the accuracy of their progress, prematurely declare completion, misattribute errors to unrelated elements, or fail to maintain a shared understanding of the current task state. This pattern relates to failures in collaborative monitoring and joint attention~\cite{tomasello2005understanding}: effective collaboration requires that partners continuously calibrate their shared model of the task state, which agents frequently fail to do. Verification failures are distributed across all four tasks, appearing most frequently in Grid Coordination (33.4\%) and Scene Comparison (27.7\%).
\emph{Example (Grid Coordination):} The verification tool reports 1 remaining difference, but both agents declare the grids ``perfectly match'' and attribute the discrepancy to ``an error in the verification system,'' prematurely declaring the task complete.

\paragraph{C2: Grounding} (17.4\%).
Agents produce ambiguous, verbose, or redundant communication that fails to establish shared understanding. This directly relates to Clark and Brennan's theory of grounding in communication~\cite{clark1991grounding}: successful collaboration requires iterative presentation and acceptance of shared referents, which agents consistently fail to achieve. Grounding failures differ from Verification (C1) in that they concern the \emph{quality of communication} rather than the \emph{accuracy of state tracking}. They are most common in Scene Comparison (42.6\%) and Tangram Reference (40.4\%), tasks where rich visual descriptions must bridge an information asymmetry.
\emph{Example (Scene Comparison):} Agent~B submits ``pink flip-flop'' as a missing item and is told it is incorrect. Agent~B resubmits the same item. Meanwhile, Agent~A describes remaining differences vaguely as ``some items near the towel,'' failing to provide the precise spatial references needed to identify the actual missing objects.

Based on the six failure patterns identified in the taxonomy, we designed a set of \emph{collaboration repair instructions} (Figure~\ref{fig:modifier}) which was appended to both agents' system prompts. Each directive translates the diagnostic insight of the corresponding pattern into actionable behavioral guidance.

\begin{figure*}[t!]
\centering
\includegraphics[width=0.95\textwidth]{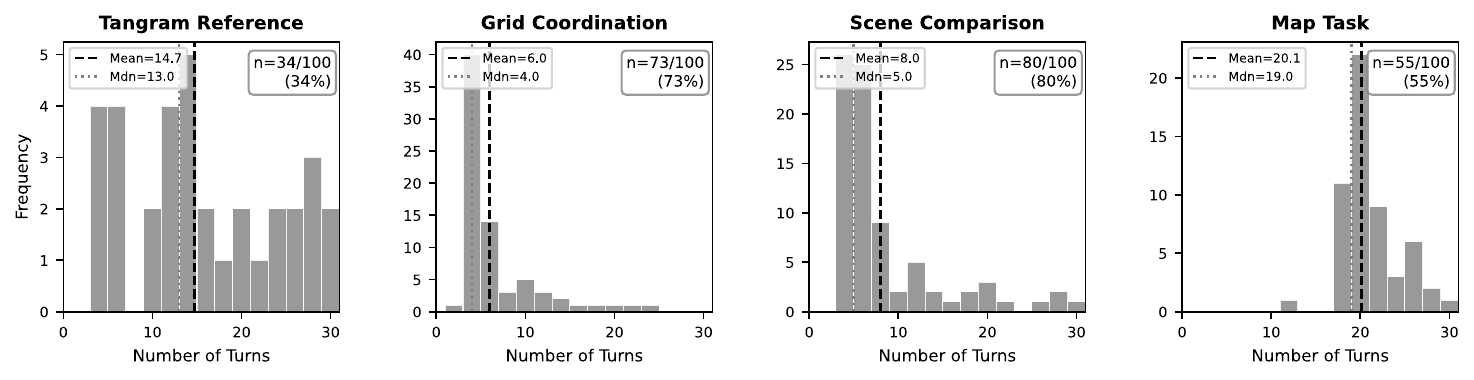}
\Description{Four histograms showing the distribution of turns to task completion for successful conversations across the four PARLEY tasks. Scene Comparison has a right-skewed distribution peaking at 3-5 turns. Grid Coordination peaks at 3-5 turns with a narrow spread. Tangram Reference shows a wider, flatter distribution spanning 3-29 turns. Map Task has completions clustered around 19-21 turns.}
\caption{\textbf{Distribution of turns to completion for successful conversations.} Each panel shows one task with mean (dashed) and median (dotted) lines. Inset labels show success count over total attempts. Grid Coordination resolves fastest (Mdn\,=\,4), while Map Task requires the most sustained coordination (Mdn\,=\,19).}
\label{fig:turn-distributions}
\end{figure*}

\section{Results}
\label{sec:results}

\begin{figure*}[t]
\centering
\includegraphics[width=0.95\textwidth]{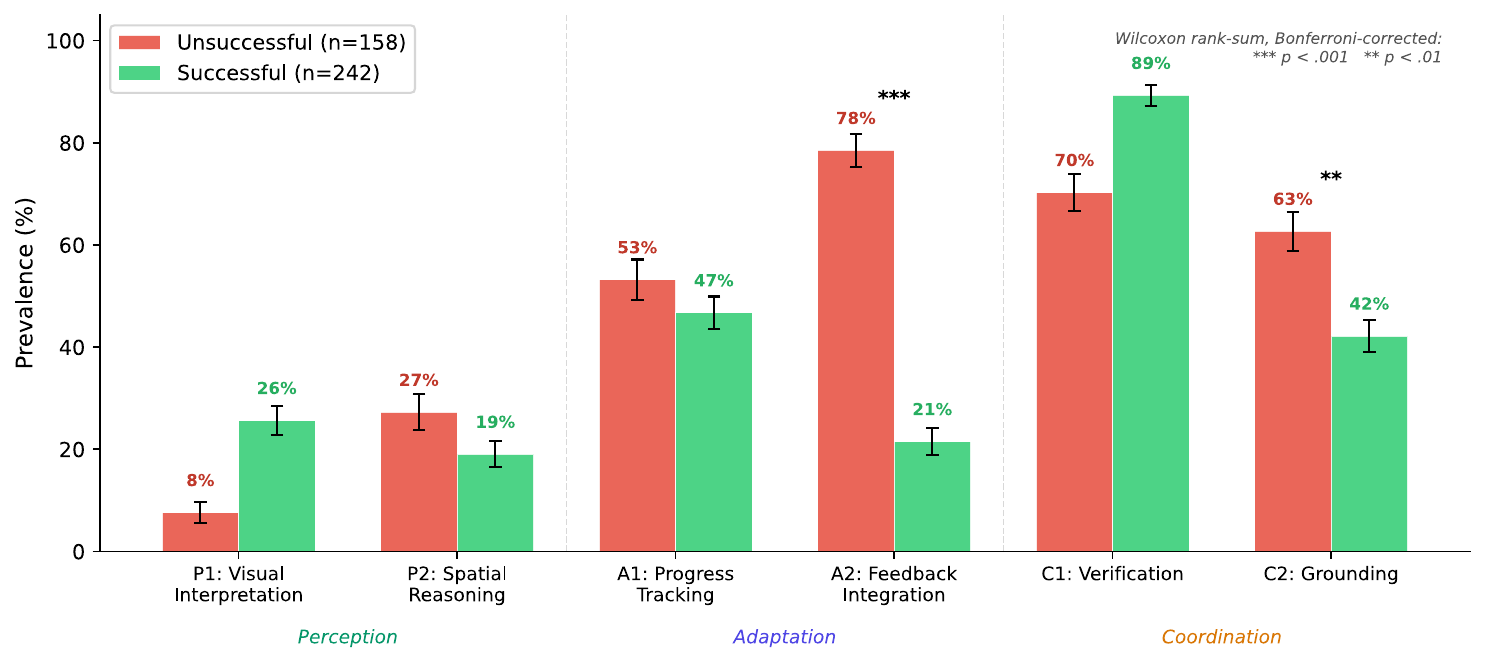}
\Description{A paired bar chart comparing the prevalence of each PAC failure pattern in successful versus incomplete conversations. A2 (Feedback Integration) shows the starkest gap, appearing in 78.5\% of failures but only 21.5\% of successes. C2 (Grounding) also shows a large gap. C1 (Verification) is paradoxically more prevalent in successes.}
\caption{\textbf{Prevalence of PAC failure patterns in successful vs.\ unsuccessful collaborations} ($N$=400). Error bars indicate standard error. A2 (Feedback Integration) is the most discriminative pattern (3.7$\times$).}
\label{fig:dual-histogram}
\end{figure*}

\subsection{How Do Agents Perform Across Collaborative Tasks?}

Across 400 controlled GPT-4o conversations (100 per task), overall success is \OverallGptFourORate\%.

Task-level analysis reveals meaningful variation in difficulty. Scene Comparison achieves the highest success (\SpotMissingGptFourORate\%), where the primary challenge lies in comparing two different visual scenes. Grid Coordination follows at \LogicGridGptFourORate\%, where both agents share the same visual information and the challenge is tool coordination (who places X vs.\ O) rather than information sharing. Map Task succeeds at \MapNavGptFourORate\%, reflecting the compounding difficulty of simultaneous visual, information, and tool asymmetry: the director must translate spatial knowledge into step-by-step navigation commands for a partner who cannot see the map. Tangram Reference achieves only \SymbolLockGptFourORate\%, requiring agents to describe abstract visual symbols with sufficient precision for a partner who cannot see them.

Performance degrades sharply with conversation length: all conversations completed in $\leq$6 turns succeed, but conversations still ongoing at turn~12 have only a 37\% chance of eventual success, and those still ongoing at turn~20 succeed just 19\% of the time. Conversations that reach the 30-turn limit never succeed, establishing a practical ``point of no return'' well before the cap. This pattern suggests that early turns are critical: if agents cannot establish effective grounding quickly, subsequent turns tend to compound rather than resolve misalignment.

Figure~\ref{fig:turn-distributions} shows the distribution of turns to completion for successful conversations across the four tasks. Grid Coordination resolves fastest when it succeeds (Mdn\,=\,4, $M$\,=\,6.0, $SD$\,=\,4.6), with a tight distribution suggesting that successful pairs quickly converge on a coordination strategy. Scene Comparison also resolves quickly (Mdn\,=\,5 turns, $M$\,=\,8.0, $SD$\,=\,6.3), likely reflecting both the relative simplicity of visual comparison and agents' ability to converge once they establish what to look for. Tangram Reference shows the widest spread (Mdn\,=\,13, $M$\,=\,14.7, $SD$\,=\,8.4), reflecting the variable difficulty of describing abstract shapes to a partner who cannot see them. Map Task demands the most turns when it does succeed (Mdn\,=\,19, $M$\,=\,20.1, $SD$\,=\,3.2), consistent with its compounding asymmetry requiring sustained coordination across many navigation steps. Overall, these distributions reflect a joint function of task difficulty and model capability: the tasks are challenging enough that a substantial fraction of conversations reach the 30-turn limit without completing, yet sufficiently tractable that successful strategies can be observed and characterized.

\subsection{Do Failure Patterns Differ Between Successful and Unsuccessful Conversations?}
\label{sec:key-findings}

Each of the 400 GPT-4o conversations is classified using the pipeline described in Sections~\ref{sec:method}--\ref{sec:taxonomy}: GPT-4o extracts failure instances from each conversation, each embedded and assigned to a cluster through hierarchical clustering. This yields per-conversation pattern counts and presence indicators for all six patterns across both successful and unsuccessful conversations, including zero counts when no instances were extracted.

Figure~\ref{fig:dual-histogram} reports pattern prevalence, the fraction of conversations containing at least one error instance of each type, computed separately for successful and unsuccessful conversations. A2 (Feedback Integration) is the most discriminative pattern, appearing in 78.5\% of failures but only 21.5\% of successes (3.7$\times$). C2 (Grounding) is also more prevalent in failures (62.7\% vs.\ 42.1\%, 1.5$\times$). The separate Wilcoxon rank-sum tests find greater raw counts in failures for A2 (Bonferroni-corrected $p < .001$) and C2 ($p = .002$). C1 (Verification) shows a paradoxical pattern: it is \emph{more prevalent} in successes (89.3\% vs.\ 70.3\%), and P1 (Visual Interpretation) is similarly concentrated in successful conversations (25.6\% vs.\ 7.6\%), mostly from Grid Coordination. This may reflect successful conversations involving more active engagement: more tool invocations, more verification attempts, more visual re-readings, which could produce more error instances of these recoverable types. P2 (Spatial Reasoning) and A1 (Progress Tracking) show moderate but non-significant directional trends toward failures in the count tests.

A LMM reveals that five of six patterns are significant indicators of failure when controlling for co-occurrence: A2 shows the largest coefficient ($\beta = 0.34$, $p < .001$), followed by P2 ($\beta = 0.26$), C2 ($\beta = 0.24$), C1 ($\beta = 0.16$), and A1 ($\beta = 0.16$, all $p < .001$). P1 is the only non-significant indicator ($\beta = 0.02$, $p = .76$). The task random-intercept variance is small ($\sigma^2_u = 0.028$, ICC\,=\,0.17), confirming that failure patterns generalize across tasks, though task identity explains a modest fraction of outcome variance.

\subsection{Can a Taxonomy-Guided Intervention Improve Collaboration?}
\label{sec:intervention}

Analysis of the intervention completion rates shows an increase in average task success from 60.5\% to 70.5\% ($+$10.0 percentage points, $p = .019$, Fisher's exact), a consistent positive trend across all four tasks (Table~\ref{tab:intervention}). Map Task shows the largest per-task gain ($+$13.0pp, from 55\% to 68\%), followed by Tangram Reference ($+$12.0pp, from 34\% to 46\%). Scene Comparison improves from 80\% to 88\% ($+$8.0pp), and Grid Coordination from 73\% to 80\% ($+$7.0pp). The consistency of improvements across tasks with very different baseline rates and collaborative demands strengthens the finding.

Among successful conversations, the intervention is also associated with shorter dialogues in some tasks. Grid Coordination completions drop from 6.0 to 4.8 turns, and Scene Comparison from 8.0 to 7.0 turns. Map Task, by contrast, shows similar turn counts (20.1 to 20.7), consistent with the sustained coordination this task demands.

\section{Discussion}

Motivated by cognitive science research on human collaboration, we evaluated GPT-4o as an agent collaboratively solving four asymmetric tasks, achieving 60.5\% overall success. Through systematic analysis of these conversations, the PAC taxonomy revealed six failure patterns that differentially characterize collaboration outcomes, and a taxonomy-guided intervention improved average task success to 70.5\%, illustrating the diagnostic utility of the framework. Here we discuss the implications of these findings.

Agent failures can be interpreted in relation to prior human-collaboration research. In human collaboration, a well-documented challenge is egocentric bias: people consistently privilege their own perspective during joint tasks, even when they know their partner holds different information~\cite{keysar2000taking, keysar2003limits, epley2004perspective}. In PARLEY's excerpts, agents sometimes show excessive compliance, capitulating rather than defending their (sometimes correct) position when partners disagree. This compliance tendency may contribute to several of the PAC failure patterns, particularly Feedback Integration (A2), where agents accept partner suggestions without critically evaluating them, and Grounding (C2), where agents confirm shared understanding prematurely rather than probing for misalignment.

\begin{table*}[t]
\centering
\footnotesize
\caption{\textbf{Intervention experiment results.} Success rates and mean turns to completion (successful conversations only) for baseline vs.\ collaboration repair instructions. $\Delta$pp\,=\,percentage-point change. Unadjusted $p$-values from two-sided Fisher's exact test (per-task and overall).}
\label{tab:intervention}
\begin{tabular*}{\textwidth}{@{\extracolsep{\fill}}lrrr|rrrrl@{}}
\toprule
 & \multicolumn{3}{c|}{\textbf{Baseline}} & \multicolumn{4}{c}{\textbf{Intervention}} & \\
\textbf{Task} & $N$ & \textbf{\%} & \textbf{Turns} & $N$ & \textbf{\%} & \textbf{Turns} & $\Delta$\textbf{pp} & $p$ \\
\midrule
Grid Coord.     & 100 & 73.0 &  6.0 & 50 & 80.0 &  4.8 & $+$7.0  & .42 \\
Map Task        & 100 & 55.0 & 20.1 & 50 & 68.0 & 20.7 & $+$13.0 & .16 \\
Scene Comp.     & 100 & 80.0 &  8.0 & 50 & 88.0 &  7.0 & $+$8.0  & .26 \\
Tangram Ref.    & 100 & 34.0 & 14.7 & 50 & 46.0 & 12.4 & $+$12.0 & .16 \\
\midrule
\textit{Overall} & 400 & 60.5 & 11.1 & 200 & 70.5 & 10.6 & $+$10.0 & .019 \\
\bottomrule
\end{tabular*}
\end{table*}
A related consideration concerns the expression of uncertainty. Human collaborators naturally signal confusion, hesitation, and doubt through hedging language, questions, and explicit statements of uncertainty, which serve as critical coordination cues for their partners~\cite{clark1991grounding, schegloff1977preference}. Some PARLEY excerpts instead show confident-sounding responses even when agents' capabilities (such as visual interpretation) may not be reliable or their environment representation is ambiguous. This is not simply a stylistic difference: the absence of uncertainty signals deprives the partner of information needed to calibrate certainty, offer help, or adjust strategy. Several of our failure patterns, particularly Grounding (C2) and Feedback Integration (A2), may be amplified by failures to surface confusion as a communicative act.

Human collaborators, when a strategy fails, naturally explore alternatives: trying different descriptions, reframing the problem, or switching roles~\cite{schegloff1977preference}. In our excerpts, agents exhibit Feedback Integration failures (A2), reapplying the same failed strategy with minor surface-level variations. This pattern is consistent with the ``passive responder'' characterization of Wu et al.~\cite{wu2025collabllm} and suggests opportunities to support explicit monitoring of strategy effectiveness. While grounding theory~\cite{clark1991grounding} documents how humans naturally adapt when communication fails, some PARLEY agents persist with failed strategies even when tool feedback explicitly indicates failure, a rigidity consistent with the Einstellung effect~\cite{luchins1942mechanization}.

These observations suggest that interventions designed for human collaboration failures (e.g., perspective-taking prompts to counter egocentric bias) may need adaptation for LLM agents~\cite{bansal2021complementary}. The PAC taxonomy provides concrete intervention points for collaborative behavior: from how agents perceive their environment, to how they adapt their strategies, to how they coordinate with their partner, and the results of our intervention experiment support the diagnostic value of this framework.

Despite the encouraging findings, this work has several limitations. For instance, we evaluate agent-agent interactions on four controlled tasks that may offer limited generalization to noisier real-life collaborative settings. While this controlled setup allowed us to study failure patterns of agentic collaboration under controlled conditions, future work may consider more realistic scenarios. Also, it is important to highlight that the PAC taxonomy has been derived from a single model (GPT-4o), which means different model families may exhibit additional failure patterns not captured by our six-pattern scheme. While we believe some of the patterns will translate to other models, we envision the same methodology can be readily applied to characterize their specific failure patterns and enable systematic cross-architecture comparison. The platform and analysis pipeline are reusable, while measured failure frequencies and intervention gains characterize GPT-4o in these tasks.

Looking ahead, this work illustrates a cycle of simulating, analyzing, and improving collaborative behavior through system-prompt updates. We explore one researcher-guided iteration, but future work could extend this process into longer, agent-directed cycles as a step toward recursive self-improvement. Agents could synthesize tasks that reflect collaboration difficulties encountered in real-world settings, use PARLEY to compare alternative strategies or instructions, and revise their own collaboration prompts based on the resulting evidence. Evaluating revisions on held-out tasks would help establish whether these cycles produce improvements beyond the situations used to generate them.

Ultimately, a broader vision for PARLEY is to draw on human collaboration research to inspire better communication and coordination among people and agents. This raises a central design question: how do the strategies that support effective agent--agent collaboration differ from those that help people understand, guide, and collaborate with an agent? Future studies could use and extend PARLEY to compare human--human, agent--agent, and human--agent interactions, examining which practices transfer across these settings and which benefit from adaptation. By connecting these forms of collaboration, PARLEY could help turn advances in agent coordination into more effective and understandable human--AI partnerships.

\section{Conclusion}

We presented PARLEY, an extensible research platform for studying LLM collaboration through four asymmetric puzzle tasks grounded in cognitive psychology. Through systematic analysis of GPT-4o conversations, we derived a data-driven taxonomy of six failure patterns, the PAC taxonomy (Perception, Adaptation, Coordination), via an extraction-embedding-clustering pipeline. We found that five of these patterns are significant indicators of collaboration failure, with Feedback Integration (A2) being the strongest. A taxonomy-guided intervention appending collaboration repair instructions improved average task success to 70.5\%. We envision a future where platforms like PARLEY enable the systematic study and iterative refinement of agent collaboration, bringing us closer to AI partners that collaborate as naturally and effectively as humans do.

\section{Generative AI Usage Disclosure}

This work leveraged generative AI tools throughout its development. GitHub Copilot (powered by Claude Opus 4.6) assisted with coding, analysis, and writing across the platform, experimental infrastructure, and manuscript, with the authors reviewing, validating, and directing each iteration. GPT-4o served both as the study's collaborative agent and as the extraction model for identifying failure instances from conversation transcripts, with cluster labels refined by the authors. The authors take full responsibility for the accuracy, originality, and intellectual contributions of this work.

\bibliographystyle{ACM-Reference-Format}
\bibliography{references}

\end{document}